\documentstyle[seceq,epsf,wrapft]{ptptex}

\def\vS{{\mib S}}
\def\vT{{\mib T}}
\def\eff{{\rm eff}}

\def\ket#1{|#1\rangle}
\def\up{\uparrow}
\def\dn{\downarrow}
\def\Up{\Uparrow}
\def\Dn{\Downarrow}

\title{
Anisotropies of the Hamiltonian and the Wave Function:
Inversion Phenomena in Quantum Spin Chains
}

\author{
Kiyomi {\sc Okamoto}\footnote{E-mail: kokamoto@stat.phys.titech.ac.jp}
}

\inst{
Department of Physics, Tokyo Institute of Technology, Tokyo 152-8551
}

\recdate{
~~~~~~~~~~~~~~~~~~~~~~~~
}

\abst{
We investigate the inversion phenomenon between the $XXZ$ anisotropies
of the Hamiltonian and the wave function in quantum spin chains,
mainly focusing on the $S=1/2$ trimerized $XXZ$ model with the
next-nearest-neighbor interactions.
We have obtained the ground-state phase diagram 
by use of the degenerate perturbation theory
and the level spectroscopy analysis of the numerical data
calculated by the Lancz\"os method.
In some parameter regions,
the spin-fluid is realized for the Ising-like anisotropy,
and the N\'eel state for the $XY$-like anisotropy,
against the ordinary situation.
}

\begin{document}

\maketitle

\section{Introduction}

In the ground state problem of the antiferromagnetic quantum spin chains
with the $XXZ$ interaction anisotropy
$\Delta \equiv J_z/J_\perp$,
the Ising-like anisotropy $\Delta>1$ is usually favorable to the N\'eel state
and the $XY$-like anisotropy $\Delta<1$ to the spin-fluid (SF) state.
This can be seen, for instance,
in the $S=1/2$ $XXZ$ chain and the $S=1$ $XXZ$ chain,
although the Haldane phase exists between the N\'eel and the SF phase
for the latter case.
Very recently we have found that the {\it inversion phenomenon}
between the anisotropies of the interaction and the wave function.\cite{oka-ichi}
Namely, the SF state is realized for the Ising-like case
and the N\'eel state for the $XY$-like case
in the $S=1/2$ distorted diamond spin chain\cite{oka-tone} with the
$XXZ$ anisotropy.
This inversion phenomenon is considered to be due to the interplay
among the frustration, the trimer nature and the $XXZ$ anisotropy of the
Hamiltonian.
In case of strong frustrations, 
the spin system tends to form singlet pairs
to avoid the energy loss by competing interactions \cite{oka-nomu1}.
But the formation of singlet pairs is incompatible with the trimer nature.
Thus, in the $XY$-like anisotropy case,
the spins turn to the $z$-direction to avoid the energy loss, 
because interaction of the $z$-direction is weaker than that of the
$xy$-direction in the $XY$-like case.
Similarly, we can explain the existence of the SF state
for the Ising-like anisotropy case.

If this physical explanation is to the point,
the inversion phenomenon will be observed in many quantum spin chain models
having the frustration, the trimer nature and the $XXZ$ anisotropy.
From this point of view,
we have investigated the $S=1/2$ trimerized $XXZ$ chain
with the next-nearest-neighbor interactions,
which may be the most fundamental model
with these three characteristics.

\section{$\mib{S =1/2}$ Trimerized ${\mib XXZ}$ Spin Chain with Next-Nearest-Neighbor Interactions}

Our model is the $S=1/2$ trimerized $XXZ$ chain
with the next-nearest-neighbor interactions (Fig. 1),
having the Hamiltonian
\begin{equation}
    H =  
    J_1 \sum_i \{h_{3i-1,3i}(\Delta) + h_{3i,3i+1}(\Delta)\}
     + J_2 \sum_i h_{3i-2,3i-1}(\Delta) 
     + J_3 \sum_j h_{j,j+2}(\Delta),
\end{equation}
where
\begin{equation}
   h_{m,n}
   \equiv S_m^x S_n^x + S_m^y S_n^y + \Delta S_m^z S_n^z.
\end{equation}
\begin{wrapfigure}{r}{6.6cm}
  \epsfxsize= 5.5 cm   
  \centerline{\epsfbox{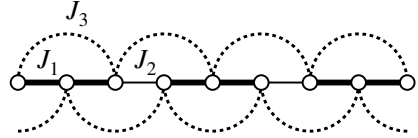}}
\caption{The $S=1/2$ trimerized $XXZ$ chain
         with the next-nearest-neighbor interactions.}
\label{fig:model}
\end{wrapfigure}
The quantity $J_1$ (thick lines) denotes the intra-trimer nearest-neighbor coupling,
$J_2$ (thin lines) the inter-trimer nearest-neighbor coupling 
and $J_3$ (broken lines) the next-nearest-neighbor coupling.
All the couplings are assumed to be positive (antiferromagnetic).
Hereafter we focus on the $J_1 \gg J_2,J_3$ case for simplicity,
where the inversion phenomenon was observed in the on
$S=1/2$ distorted diamond chain with $XXZ$ anisotropy.

Let us discuss the ground state of our model by use
of the degenerate perturbation theory.
First we consider the 3-spin problem of $\vS_{3i-1},\vS_{3i}$ and $\vS_{3i+1}$.
As far as $J_1 \gg J_3$,
the ground states of the 3-spin cluster are
\begin{eqnarray}
    &&\ket{\Up_i}
    \equiv {1 \over A} (\ket{\up\up\dn} - a\ket{\up\dn\up} + \ket{\dn\up\up}) 
    \\
    &&\ket{\Dn_i}
    \equiv {1 \over A} (\ket{\up\up\dn} - a\ket{\up\dn\up} + \ket{\dn\up\up})
\end{eqnarray}
where $\ket{\up\up\dn}$ means $\ket{\up_{3i-1}\up_{3i}\dn_{3i+1}}$
for instance,
$A=\sqrt{2+a^2}$ and
\begin{equation}
    a = {- 3\tilde J_3 + \Delta + r \over 2 + \tilde J_3^2 + \tilde J_3 \Delta - J_3 r},~~~~~
    r = \sqrt{8 + (\tilde J_3 + \Delta)^2}
\end{equation}
where $\tilde J_3 \equiv J_3/J_1$.
As far as $J_1 \gg J_2,J_3$,
we can restrict ourselves to these two states, $\ket{\Up_j}$ and $\ket{\Dn_j}$, 
for the $j$th trimer,
neglecting other 6 states.
For convenience we consider $\ket{\Up_j}$ and $\ket{\Dn_j}$ as 
the up-spin and down-spin states of the pseudo-spin $\vT_j$, respectively.
The interactions between the trimers are expressed as the interactions
between pseudo-spins.
For instance, we have
\begin{equation}
    {J_3 \over 2}
    \left( S_{3j}^+ S_{3j+2}^- + S_{3j}^- S_{3j+2}^+ \right)
    \Rightarrow
    - {2aJ_3 \over (a^2+2)^2}
    \left(T_j^+ T_{j+1}^- + T_j^- T_{j+1}^+\right)
\end{equation}
In the lowest order with respect to $J_2$ and $J_3$,
a straightforward calculation  leads to
\begin{equation}
    H_{\rm eff}
    = \sum_j \left\{
        J_{\rm eff}^\perp \left(T_j^x T_{j+1}^x + T_j^y T_{j+1}^y \right)
        + J_{\rm eff}^z T_j^z T_{j+1}^z \right\}
\end{equation}
where
\begin{eqnarray}
    &&J_{\rm eff}^\perp
    = {a \over (a^2+2)^2}\left\{ 4aJ_2 - 8J_3\right\} \\
    &&J_{\rm eff}^z
    = {a \over (a^2+2)^2}\left\{ a^2(a^2-2)J_2 - 2(a^2-2)^2J_3) \right\}
\end{eqnarray}
In the effective theory of the lowest order with respect to
$J_2/J_1$ and $J_3/J_3$,
we can let $J_3=0$ in $a$, that is
\begin{equation}
    a = {\Delta + \sqrt{\Delta^2 + 8} \over 2}
\end{equation}

When $\Delta=1$, we see $J_{\rm eff}^\perp = J_{\rm eff}^z = (4/9)(J_2-J_3)$.
Then the ground state of $H_{\rm eff}$ is either the ferromagnetic state
or the spin-fluid state depending on whether $J_2<J_3$ or $J_2>J_3$.
The ferromagnetic state of the $\vT$-system corresponds to
the ferrimagnetic state of the $\vS$-system with the magnetization of $M_{\rm s}/3$,
where $M_{\rm s}$ is the saturation magnetization.

In general case ($\Delta \ne 1$),
the ground state of $H_{\rm eff}$ is known from $J_{\rm eff}^\perp$ and $J_{\rm eff}^z$
as 
\begin{equation}
  \matrix{
    J_\eff^z <0 \hbox{ and } |J_\eff^z| > |J_\eff^\perp|
    &~~\Rightarrow~~
    &\hbox{ferromagnetic state} \hfill\cr
    |J_\eff^z| < |J_\eff^\perp| \hfill
    &~~\Rightarrow~~
    &\hbox{spin-fluid state} \hfill\cr
    J_\eff^z > |J_\eff^\perp| \hfill
    &~~\Rightarrow~~
    &\hbox{N\'eel state} \hfill
  }
  \label{eq:condition}
\end{equation}
which leads to
\begin{equation}
  \matrix{
                             &&\hbox{$XY$-like}   &\hbox{Ising-like} \cr
    J_3 < c_1 J_2            &&\hbox{SF}          &\hbox{N\'eel  }   \cr
    c_1 J_2 < J_3 < c_2 J_2  &&\hbox{N\'eel}      &\hbox{SF} \cr
    J_3 > c_2 J_2            &&\hbox{SF}          &\hbox{ferro}
    }
    \label{eq:condition2}
\end{equation}
where
\begin{equation}
    c_1 = {a^2+2a+2 \over 2a(a+2)}~~~~~~~
    c_2 = {1 \over 2}{a(a+2)(a^2-2a+2) \over a^4-4a^2+8}
    \label{eq:c1c2}
\end{equation}
We note that 
\begin{equation}
    c_1 \to 5/8~~~~~c_2 \to 1~~~~~(\hbox{as }\Delta \to 1)
\end{equation}
The phase diagrams for the $\Delta=1,\ 0.5\mbox{~and~}2.5$ cases 
near the truncation point $J_2 = J_3 =0$ are shown
in Figs. 2-4,
where $\tilde J_2 \equiv J_2/J_1,\ \tilde J_3 \equiv J_3/J_1$.

\begin{figure}[htb]
    \parbox{.3\textwidth}{
        \epsfxsize= .3\textwidth   
        \centerline{\epsfbox{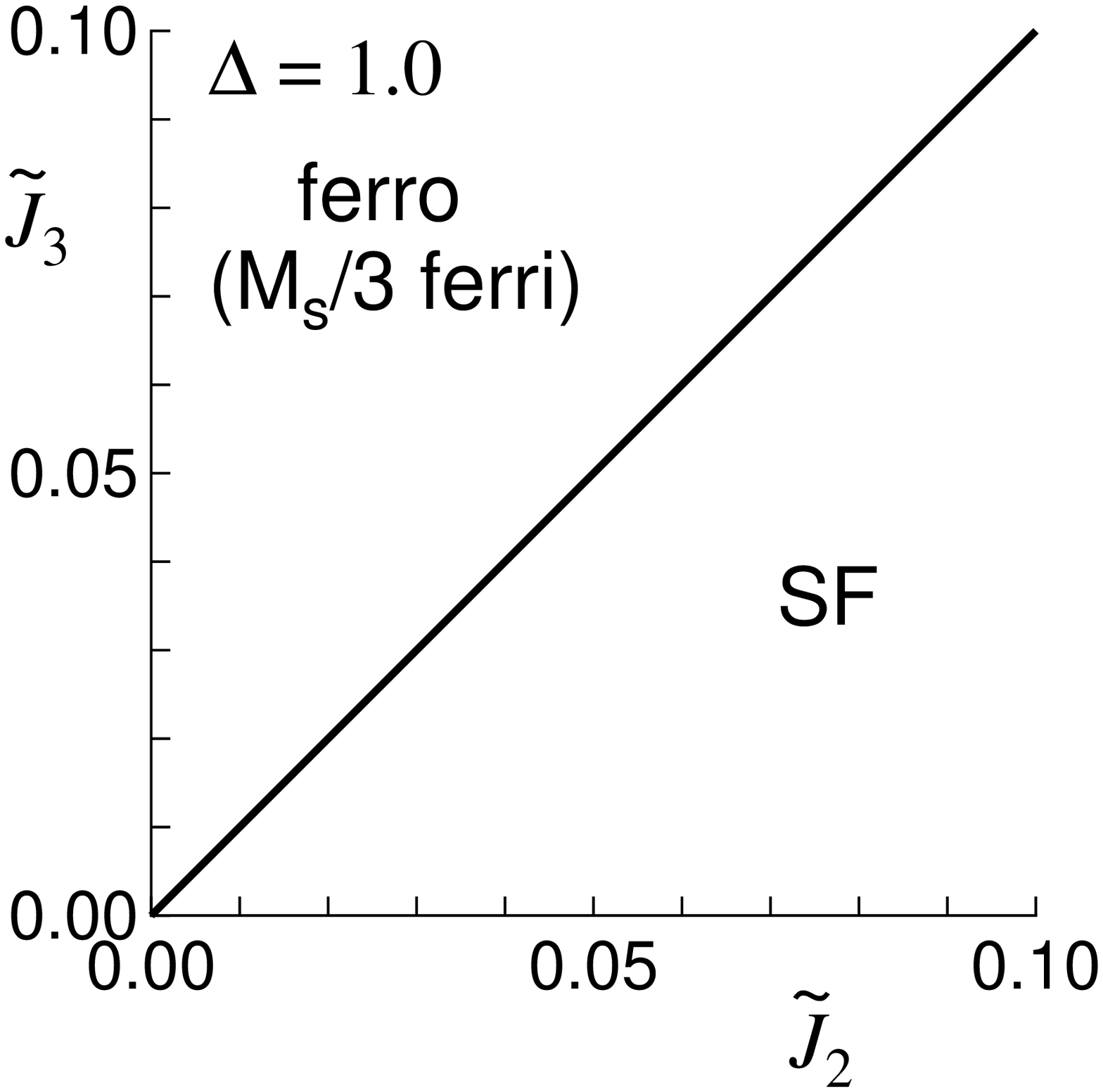}}
        \caption{Phase diagram for the $\Delta=1$ case.}
        \label{fig2}}
    \hspace{4mm}
    \parbox{.3\textwidth}{
        \epsfxsize= .3\textwidth   
        \centerline{\epsfbox{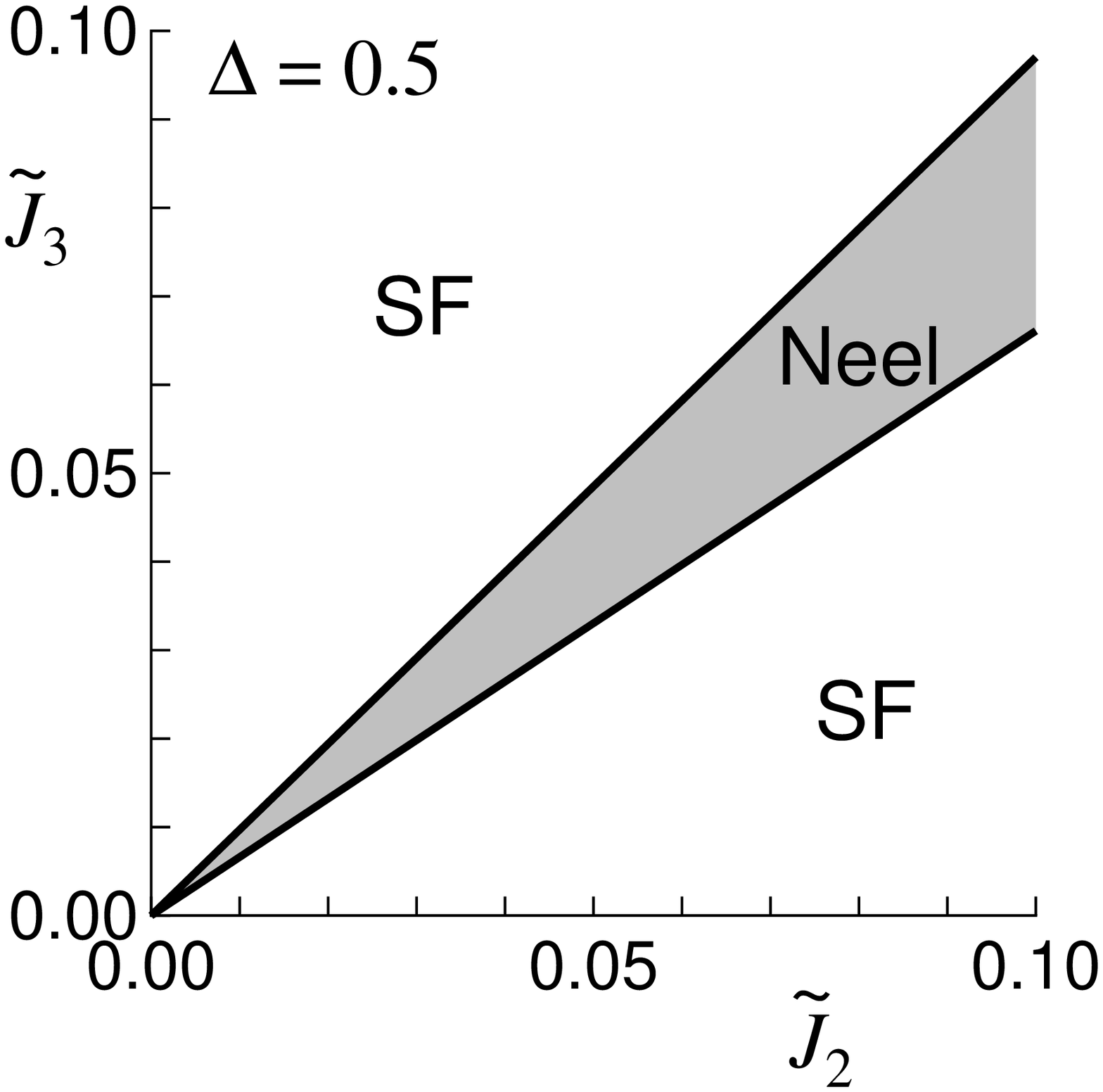}}
        \caption{Phase diagram for the $\Delta=0.5$ case.}
        \label{fig3}}
    \hspace{4mm}
    \parbox{.3\textwidth}{
        \epsfxsize= .3\textwidth   
        \centerline{\epsfbox{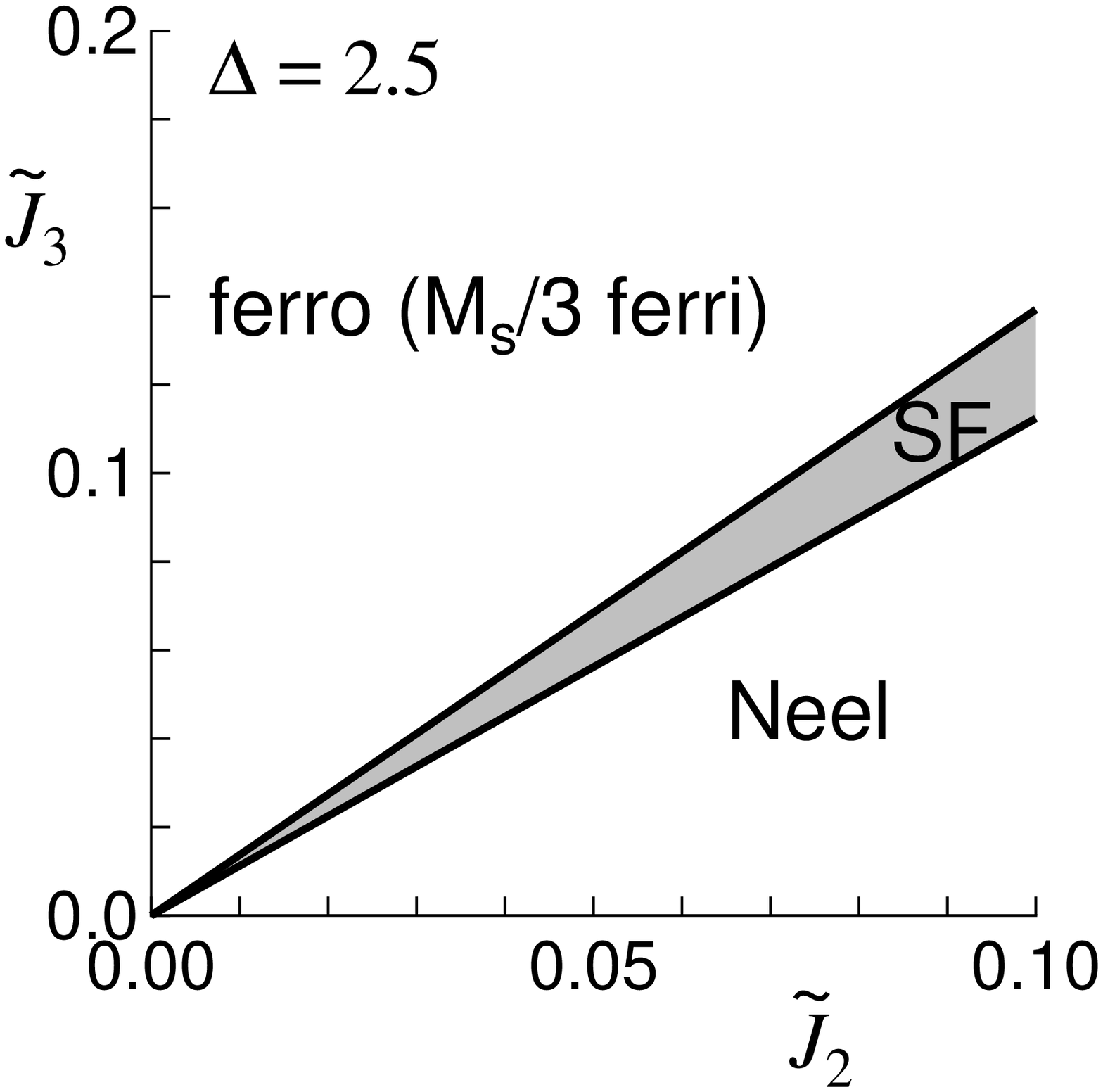}}
        \caption{Phase diagram for the $\Delta=2.5$ case.}
        \label{fig4}}
\end{figure}

In the Ising limit $\Delta \to \infty$,
we see $c_1 = c_2 = 1/2$,  resulting in the vanishing of the
spin-fluid state in Fig. 4.
The remarkable nature of Fig. 3 is the existence of the
N\'eel region,
although the interaction anisotropy is $XY$-like. 
Similarly, the SF state is realized in Fig. 4
in spite of the Ising-like anisotropy.
These are the {\it inversion phases}.

We have also performed the numerical diagonalization by use of the
Lancz\"os method and analyzed the numerical data by the
level spectroscopy method.\cite{LS,oka-nomu1,oka-nomu2,oka-nomu3}
For instance, when $\tilde J_2 = 0.01$ and $\Delta = 0.5$,
the critical values of the SF-N\'eel transitions are
$(\tilde J_3)_{\rm cr1} = 0.006610$ and $(\tilde J_3)_{\rm cr2} = 0.009696$,
showing good agreements with the theoretical values
from Eqs.(\ref{eq:condition2}) and (\ref{eq:c1c2})
$(\tilde J_3)_{\rm cr1} = 0.006609$ and $(\tilde J_3)_{\rm cr2} = 0.009703$,
respectively.
For the $\tilde J_2 = 0.01$ and $\Delta = 2.5$ case,
we have obtained
$(\tilde J_3)_{\rm cr1} = 0.005623$ (N\'eel-SF) and 
$(\tilde J_3)_{\rm cr2} = 0.006863$ (SF-ferrimagnetic),
also having good agreements with the theoretical values
$(\tilde J_3)_{\rm cr1} = 0.006609$ and $(\tilde J_3)_{\rm cr2} = 0.006850$,
respectively.

\section{Concluding Remarks}

We have shown that there exist inversion regions in the
$S=1/2$ trimerized $XXZ$ chain with next-nearest-neighbor interactions
by use of the degenerate perturbation theory as well as the level spectroscopy
analysis\cite{LS} of the numerical data obtained by the Lancz\"os method.
This strongly suggests that the inversion phenomenon is popular to
the $=1/2$ $XXZ$ models with the frustration and the trimer nature.
In fact, this phenomenon has been also found in the $S=1/2$
frustrated 3-leg ladder with the $XXZ$ anisotropy.\cite{oka-saka}

In this paper we have restricted ourselves to the $J_1 \gg J_2,J_3$ case
for simplicity.
In case of larger $J_3$, the dimer ground state is expected.
In fact, when $J_2 = J_1$, our model is reduced to the uniform $XXZ$ chain
with next-nearest-neighbor interactions, 
which shows the SF-dimer or N\'eel-dimer
transitions.\cite{oka-nomu1,oka-nomu2,oka-nomu3}
The full phase diagram of the present model will be reported elsewhere.


\end{document}